\makeatletter
\input{filecontents.sty}
\makeatother

\documentclass[aps,prl,reprint,showpacs,secnumarabic,superscriptaddress]{preprint}

\renewcommand\documentclass[2][]{}
\documentclass[aps,prl,reprint,showpacs,floatfix,superscriptaddress]{revtex4-1}
\usepackage[T1]{fontenc}
\usepackage{amsmath}
\usepackage{txfonts}
\usepackage{textcomp}
\usepackage{microtype}
\usepackage{graphicx}
\usepackage[
breaklinks=true,
pdfauthor={Sven Ahrens, Heiko Bauke, Christoph H. Keitel, and Carsten M{\"u}ller},
pdftitle={Spin dynamics in the Kapitza-Dirac effect}
]{hyperref}
\usepackage{color}
\usepackage{braket}
\newcommand*{\I}{\mathrm{i}}
\newcommand*{\e}{\textrm{e}}

\graphicspath{{figures/}}

\renewcommand*{\vec}[1]{\boldsymbol{#1}}

\begin{document}

\title{Spin dynamics in the Kapitza-Dirac effect}

\author{Sven Ahrens}
\author{Heiko Bauke}
\email{heiko.bauke@mpi-hd.mpg.de}
\author{Christoph H. Keitel}
\affiliation{Max-Planck-Institut f{\"u}r Kernphysik,
  Saupfercheckweg~1, 69117~Heidelberg, Germany}
\author{Carsten M{\"u}ller}
\email{mueller@tp1.uni-duesseldorf.de}
\affiliation{Max-Planck-Institut f{\"u}r Kernphysik,
  Saupfercheckweg~1, 69117~Heidelberg, Germany}
\affiliation{Institut f{\"u}r Theoretische Physik~I,
  Heinrich-Heine-Universit{\"a}t D{\"u}sseldorf, Universit{\"a}tsstra{\ss}e~1,
  40225~D{\"u}sseldorf, Germany}

\date{\today}

\begin{abstract}
  Electron spin dynamics in Kapitza-Dirac scattering from a standing
  laser wave of high frequency and high intensity is studied. We develop
  a fully relativistic quantum theory of the electron motion based on
  the time-dependent Dirac equation.  Distinct spin dynamics, with Rabi
  oscillations and complete spin-flip transitions, is demonstrated for
  Kapitza-Dirac scattering involving three photons in a parameter regime
  accessible to future high-power \mbox{X-ray} laser sources.  The Rabi
  frequency and, thus, the diffraction pattern is shown to depend
  crucially on the spin degree of freedom.
\end{abstract}


\maketitle

\paragraph{Introduction}

The diffraction of an electron beam from a standing wave of light is
referred to as the Kapitza-Dirac effect
\cite{Kapitza:Dirac:1933:effect,Batelaan:2007:Illuminating_Kapitza-Dirac}.
This process points out the quantum wave nature of the electron and may
be considered as an analogue of the optical diffraction of light on a
grating, but with the roles of light and matter interchanged. Predicted
already in 1933, a clear experimental confirmation of the Kapitza-Dirac
effect as originally proposed has been achieved only recently
\cite{Freimund:etal:2001:Observation_Kapitza-Dirac,*Freimund:Batelaan:2002:Bragg}.
It has stimulated renewed theoretical interest in the process
\cite{Smirnova:etal:2004:two_color_Kapitza-Dirac,*Li:etal:2004:Theory_Kapitza-Dirac,*Sancho:2010:two-particle_Kapitza-Dirac},
advancing earlier treatments
\cite{Fedorov:1967:Kapitza_Dirac,Gush:1971:Electron_Scattering,*Fedorov:1974:Stimulated_scattering}.
A related successful experiment observed the (classical) scattering of
electrons in a standing laser wave
\cite{Bucksbaum:etal:1988:Kapitza-Dirac_Effect}. Both experiments
operated at moderate light intensities between
$10^{9}$\,W/cm\textsuperscript{2} and $10^{14}$\,W/cm\textsuperscript{2}
and near optical wavelengths. The existing theoretical studies,
accordingly, have treated the electron quantum dynamics
nonrelativistically. The nonrelativistic Kapitza-Dirac effect has been
observed experimentally also using atomic beams
\cite{Gould:etal:1986:Diffraction_of_atoms,*Adams:etal:1994:Atom_optics,*Freyberger:etal:1999:Atom_Optics}.

The current development of novel light sources, envisaged to provide
field intensities in excess of $10^{20}\,\text{W/cm\textsuperscript{2}}$
and field frequencies in the hard \mbox{X-ray} domain
\cite{Altarell:etal:2007:XFEL,McNeil:Thompson:2010:XFEL,*Emma:etal:2010:First_lasing,*Mourou:Tajima:2011:More_intense},
raises the question as to how the Kapitza-Dirac effect is modified in
this hitherto unexplored parameter regime. This calls for a fully
relativistic treatment of the process within the Dirac theory, valid at
high electron energies and, in particular, accounting for the electron
spin.  Relativistic considerations of quantum mechanical electron
scattering in two counterpropagating light waves were provided, but
assuming nonequal laser frequencies and disregarding the electron spin
degree of freedom
\cite{Haroutunian:Avetissian:1975:analogue_KGE,*Federov:McIver:1980:compton_scattering}.
Spin signatures in the Kapitza-Dirac effect have so far been examined
within the nonrelativistic framework of the Pauli equation
\cite{Rosenberg:2004:Extended_theory} and by solving the classical
equations of motion \cite{Freimund:Batelaan:2003:Stern-Gerlach}. Both
studies found negligibly small spin effects.  We note that the influence
of the electron spin has also been investigated with respect to
free-electron motion
\cite{Krekora:Su:Grobe:2001:spin_of_relativistic_electrons,*Walser:Urbach:Hatsagortsyan:Hu:Keitel:2002:spin_and_radiation_in_intense_laser_fields},
bound-electron dynamics \cite{Hu:Keitel:1999:Spin_Signatures}, atomic
photoionization
\cite{Faisal:Bhattacharyya:2004:helium_double_ionization_spin_assymetry},
and Compton and Mott
\cite{Szymanowski:etal:1998:laser_scattering,*Ivanov:etal:2004:polarization,*Ehlotzky:etal:2009:Fundamental_processes}
scattering in strong plane-wave laser fields.

In this Letter, we present a fully relativistic consideration of the
Kapitza-Dirac effect within the framework of Dirac theory. We focus on
the relevance of spin-flip transitions when the process occurs in
\mbox{X-ray} laser fields of high intensity.  We demonstrate that under
suitable conditions, the diffraction probability depends considerably on
the electron spin. Thus, the spin degree of freedom significantly
influences the quantum mechanical scattering dynamics.

\begin{figure}[b]
  \centering%
  \includegraphics[width=0.95\linewidth]{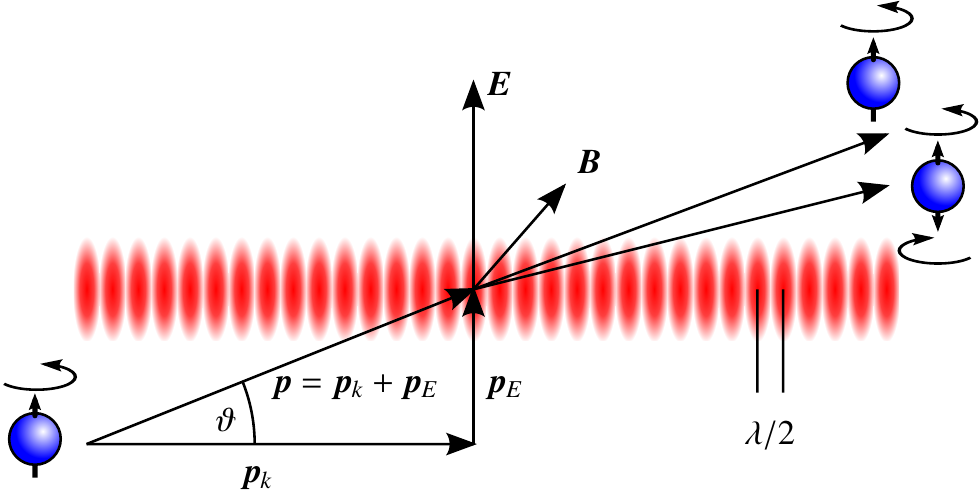}
  \caption{(Color online.)  Schematic setup. An electron with momentum
    $\vec p$ is incident at an angle $\vartheta$ on a linearly polarized
    standing laser wave with electric and magnetic components $\vec E$
    and $\vec B$. The momentum $\vec p$ has components $\vec p_E$ and
    $\vec p_k$ along the laser's electric field $\vec E$ and wave vector
    $\vec k$, respectively. The electron is initially spin-polarized
    along the electric field component. After Kapitza-Dirac scattering,
    parts of the electron wave packet may have flipped their spin
    orientation.}
  \label{fig:kde-setup}
\end{figure}

\paragraph{Relativistic theory}

We consider a quantum electron wave packet in a standing linearly
polarized light wave (see Fig.~\ref{fig:kde-setup}) with maximal
electric field strength $\vec E$, wave vector $\vec k$, and wavelength
$\lambda=2\pi/|\vec k|=2\pi/k$, respectively. The laser is modeled by
the vector potential
\begin{equation}
  \vec A(\vec x,t) = 
  -\frac{\vec E}{k} \cos(\vec k \cdot \vec x) \sin(c k t) w(t)
  \label{eq:A}
\end{equation}
introducing the speed of light $c$ and the temporal envelope function
$w(t)$.  The relativistic quantum dynamics of an electron with mass $m$
and charge $-e$ is governed by the Dirac equation \footnote{In this
  article, we use Gauss units and define $\hbar=1$.}
\begin{equation}
  \label{eq:Dirac}
  \I \frac{\partial }{\partial t} \psi(\vec x,t)
  = \left[
    c \left( 
      -\I\nabla + \frac{e}{c}\vec A (\vec x, t)
    \right)\cdot \vec \alpha + \beta m c^2 
  \right]\psi(\vec x,t)\,.
\end{equation}
In (\ref{eq:Dirac}), we have introduced the vector
$\vec\alpha=(\alpha_1,\alpha_2,\alpha_3)$, where $\alpha_i$ and $\beta$
are the Dirac matrices in standard representation
\cite{Thaller:1992:Dirac}.  The monochromatic light wave (\ref{eq:A})
allows us to decompose the wave function $\psi(\vec x,t)$ into a
discrete set of plane waves, viz.,
\begin{equation}
  \psi(\vec x,t) = 
  \sum_{n,\zeta} c_n^{\zeta}(t) \psi_{n,\vec p}^\zeta (\vec
  x)\,,\quad 
  \psi_{n,\vec p}^\zeta (\vec x)=u_{n,\vec p}^{\zeta} \e^{\I
    (\vec p + n \vec k)\cdot \vec x}\,.
  \label{eq:plane-wave-ansatz} 
\end{equation}
The function $\psi_{n,\vec p}^\zeta(\vec x)$ denotes a free particle
Dirac eigenfunction of momentum $\vec p+n\vec k$ ($n=0,\pm1, \pm2,\dots$).
The index $\zeta \in \mbox{$\{ {+}\uparrow , {+}\downarrow, {-}\uparrow,
  {-}\downarrow \}$}$ labels the sign of the energy and the spin
projection along the laser electric field vector.  Taking advantage of
the basis functions' orthonormality the ansatz
(\ref{eq:plane-wave-ansatz}) yields
\begin{multline}
  \I \dot c_n^\gamma(t) = \I \braket{\psi_{n,\vec p}^\gamma|\dot \psi}
  =
  \epsilon^\gamma \mathcal{E}(\vec p + n\vec k) c_n^\gamma(t) \\
  - 
  \frac{w(t) e \sin(c k t)}{2 k} \sum_{\zeta} 
  \braket{u_{n,\vec p}^\gamma|\vec E\cdot \vec \alpha|u_{n-1,\vec p}^{\zeta}}
  c_{n-1}^{\zeta}(t) \\
  -
  \frac{w(t) e \sin(c k t)}{2 k} \sum_{\zeta} 
  \braket{u_{n,\vec p}^\gamma|\vec E\cdot \vec \alpha|u_{n+1,\vec p}^{\zeta}} 
  c_{n+1}^{\zeta}(t) \,,
  \label{eq:FSME-dirac-equation}
\end{multline}
where we have introduced the relativistic energy momentum dispersion
relation $\mathcal{E}(\vec p)=\sqrt{m^2 c^4 + c^2 {\vec p}^2}$ and the
signum $\epsilon^\gamma$, which is $1$ for $\gamma \in \{ {+}\uparrow ,
{+}\downarrow \}$ and ${-}1$ for $\gamma \in \{ {-}\uparrow ,
{-}\downarrow \}$.

\paragraph{Generalized Bragg condition} 

The elastic scattering of an electron on a standing light wave may be
characterized by a Bragg condition \cite{Kapitza:Dirac:1933:effect}
provided that the ponderomotive energy of the electron is small
(so-called Bragg regime)
\cite{Fedorov:1967:Kapitza_Dirac,Batelaan:2000:Kapitza-Dirac}.  This
Bragg condition may be generalized to an inelastic process of absorbing
and emitting an arbitrary number of photons by utilizing momentum
conservation
$
\vec p' = \vec p + (n_r-n_l)\vec k
$
and energy conservation
$
  \mathcal{E}(\vec p') = \mathcal{E}(\vec p) + (n_r + n_l)ck \,,
$
where $\vec p$ and $\vec p'$ are the initial and final electron
momenta. The integers $n_r$ and $n_l$ denote the net numbers of photons
exchanged with the right- and left-traveling laser waves, respectively,
with positive (negative) values indicating photon absorption (emission).
The momentum and energy conservation laws
yield the relativistic generalized Bragg condition
\begin{multline} 
  \label{eq:gen_Bragg}
  \frac{\cos\vartheta}{\lambda_p} = \\
  -\frac{n_r-n_l}{2\lambda} + 
  \frac{n_r-n_l}{|n_r-n_l|}
  \frac{n_r+n_l}2\sqrt{
    \frac{1}{\lambda^2} -
    \frac{1}{n_r n_l}\left(
      \frac{\sin^2\vartheta}{\lambda^2_p} +
      \frac{1}{\lambda_C^2}
    \right)
  }
\end{multline}
by introducing the angle $\vartheta$ (see Fig.~\ref{fig:kde-setup}), the
de~Broglie wavelength $\lambda_p=2\pi/|\vec p|$, and the Compton
wavelength $\lambda_C=2\pi/(mc)$.  To be consistent with the
nonrelativistic limit $n_r$ and $n_l$ must have opposite signs.
Equation (\ref{eq:gen_Bragg}) reduces to the Bragg condition of the
\mbox{two-photon} Kapitza-Dirac effect
\cite{Kapitza:Dirac:1933:effect,Batelaan:2000:Kapitza-Dirac} by setting
$n_r=-n_l=1$.

From Eq.~(\ref{eq:gen_Bragg}), it follows that for inelastic processes
\mbox{($n_r+n_l\ne0$)} either the initial electron momentum $\vec p$ or
the laser photon momentum $k$ must be of the order of $m c$, i.\,e.,
relativistic, except we allow for a very large number of interacting
photons. Thus, an analysis of inelastic Kapitza-Dirac scattering demands
a relativistic treatment by the Dirac equation.

\paragraph{A setup for spin-sensitive Kapitza-Dirac scattering}

Equation~(\ref{eq:FSME-dirac-equation}) couples momentum components
having momenta that differ by $\pm \vec k$ and equal or opposite spin
orientation. An explicit calculation of
$\braket{u_{n,\vec p}^\gamma|\vec E\cdot\vec \alpha|u_{n\pm 1,\vec p}^{\zeta}}$
reveals that if $\vec p$ and $\vec E$ are orthogonal then
$c_n^\gamma(t)$ couples only to components $c_{n\pm 1}^\zeta(t)$ having
opposite spin orientation. Therefore, a distinct spin dynamics may be
expected for Kapitza-Dirac scattering with an odd number of photons
provided that the initial electron momentum $\vec p$ is almost
orthogonal to the electric field $\vec E$. Thus, we focus on a
\mbox{three-photon} Kapitza-Dirac effect, i.\,e., $n_r=2$, $n_l=-1$, in
the subsequent sections.

The condition (\ref{eq:gen_Bragg}) allows us to determine the initial
electron momentum $\vec p$ and the laser wave number $k$ for a resonant
\mbox{three-photon} Kapitza-Dirac effect.  The impinging electron is
modeled by a plane wave; thus, the initial condition
$c_0^{+\uparrow}(0)=1$ and $c_n^\zeta(0)=0$ else will be applied for the
remainder of the Letter.  We solve the Dirac
equation~(\ref{eq:FSME-dirac-equation}) until time $T$ to compute the
transition amplitudes $c_n^{\zeta}(T)$.

\begin{figure}[t]
  \centering
  \includegraphics[scale=0.45]{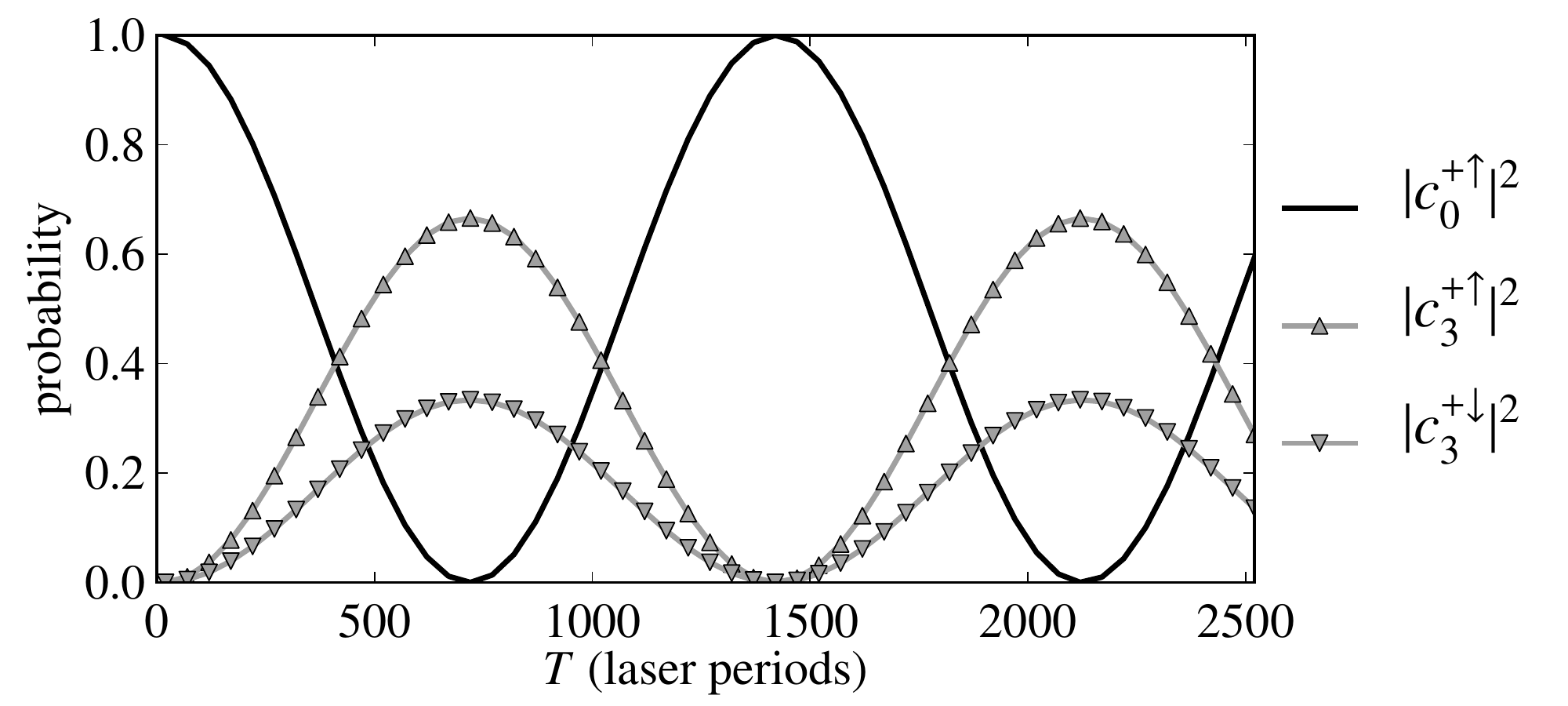}
  \caption{Rabi oscillations of the spin resolved diffraction
    probabilities as a function of the interaction time $T$ for a
    \mbox{three-photon} Kapitza-Dirac effect.  Starting from a pure
    spin-up state the electron is either diffracted with probability
    $|c_3^{+\uparrow}(T)|^2 + |c_3^{+\downarrow}(T)|^2$ or passes the
    laser beam without momentum transfer with probability
    $|c_0^{+\uparrow}(T)|^2$.  Laser parameters of this simulation
    correspond to two counterpropagating \mbox{X-ray} laser beams with a
    peak intensity of $2.0\times10^{23}\textrm{W}/\textrm{cm}^2$ each
    and a photon energy of $3.1\,\textrm{keV}$. The electron impinges at
    an angle of inclination of $\vartheta=0.4$\textdegree\ and a
    momentum of $176\,\textrm{keV}/c$, in order to fulfill the Bragg
    condition~\eqref{eq:gen_Bragg}.}
  \label{fig:spin_flip}
\end{figure}

\paragraph{Spin-flip probability and Rabi frequency} 

Figure~\ref{fig:spin_flip} shows the final occupation probabilities
$|c_n^{\zeta}(T)|^2$ after laser-electron interaction, which have been
calculated by numerical propagation of the Dirac
equation~(\ref{eq:FSME-dirac-equation}) by a combination of the explicit
and the implicit Euler method \cite{*[{Other numerical schemes for
    solving the time-dependent Dirac equation have been developed
    recently in }] [{; }]
  Braun:Su:Grobe:1999:FFT_split_operator,*Selsto:Lindroth:Bengtsson:2009:hydrogen_dirac_solution,*Mocken:2008:FFT_split_operator,*Bauke:Keitel:2011:Accelerating,*FillionGourdeau:Lorin:Bandrauk:2012:solution_dirac_equation}.
The laser profile was modeled by an envelope function $w(t)$ that starts
with a $\sin^2$-shaped turn-on ramp of 10 laser cycles and finishes with
a $\sin^2$-shaped turn-off ramp of 10 laser cycles having a flat plateau
in between.  The experimental setup was chosen to meet the Bragg
condition~(\ref{eq:gen_Bragg}) for a three-photon process.  As
illustrated in Fig.~\ref{fig:spin_flip}, only \mbox{zeroth-order} and
\mbox{third-order} modes are nonzero after interaction.  The occupation
probabilities oscillate in Rabi cycles of frequency $\Omega_R$
\begin{subequations}
  \label{eq:transtion-propability}
  \begin{align}
    \label{eq:0-mode-transtion-propability}
    |c_0^{+\uparrow}(T)|^2 & =
    \cos^2\left({\Omega_R T}/{2}\right) \,, \\
    \label{eq:3-mode-transtion-propability}
    |c_3^{+\uparrow}(T)|^2 + |c_3^{+\downarrow}(T)|^2 & =
    \sin^2\left({\Omega_R T}/{2}\right) \,,
  \end{align}
\end{subequations}
similarly to the \mbox{two-photon} Kapitza-Dirac effect
\cite{Batelaan:2007:Illuminating_Kapitza-Dirac} and the Kapitza-Dirac
effect in atomic beams
\cite{Gould:etal:1986:Diffraction_of_atoms,*Adams:etal:1994:Atom_optics,*Freyberger:etal:1999:Atom_Optics}.
In the present case, however, the scattered portion of the electron wave
packet consists of two parts which are distinguished by opposite spin
orientations. The period of the Rabi oscillations is $2\pi / \Omega_R =
1.9\,\text{fs}$ for parameters of Fig.~\ref{fig:spin_flip}.

For short times with $\Omega_RT\ll 1$, the Dirac
equation~(\ref{eq:FSME-dirac-equation}) can also be solved analytically
via time-dependent perturbation theory.  This yields for parameters
compatible with the three-photon Bragg condition
(\ref{eq:gen_Bragg}) 
the probabilities
\begin{subequations}
  \label{eq:c_3_2}
  \begin{align}
    \label{eq:c_3_2_up}
    |c_{3}^{+\uparrow}(T)|^2 & = 
    \left(
      \frac{1}{2}
      \Omega_{0} T
    \right)^2
    \left(\frac{5}{\sqrt{2}}\frac{|\vec p_E|}{k}\right)^2 
    \,,\\
    \label{eq:c_3_2_down}
    |c_{3}^{+\downarrow}(T)|^2 & = 
    \left(
      \frac{1}{2}
      \Omega_{0} T
    \right)^2 \,,
  \end{align}
\end{subequations}
with $\Omega_0={e^3|\vec E|^3}/({24m^3c^5k^2})$. The $|\vec E|^3$
dependence clearly indicates the \mbox{three-photon} nature of the
transition.  From Eq.~(\ref{eq:c_3_2}), we can derive the spin-flip
probability $P_\mathrm{flip}$ within the scattered portion of the
electron wave packet
\begin{equation}
  \label{eq:spin-flip-probability}
  P_\mathrm{flip} \equiv
  \frac{|c_{3}^{+\downarrow}(T)|^2}
  {|c_{3}^{+\uparrow}(T)|^2+|c_{3}^{+\downarrow}(T)|^2} = 
  \frac{1}{
    \frac{25}{2}\left(
      \frac{|\vec p_E|}{k}
    \right)^2 + 1} \,.
\end{equation}
The Rabi frequency may be derived by identifying (\ref{eq:c_3_2}) with
the Taylor expansion of (\ref{eq:3-mode-transtion-propability}) for
short times $T$, resulting in
\begin{equation}
  \label{eq:Rabi_DE}
  \Omega_R = \Omega_0
  \sqrt{\frac{25}{2}\left(
      \frac{|\vec p_E|}{k}
    \right)^2 + 1}\,.
\end{equation}

\begin{figure}[t]
  \centering
  \includegraphics[scale=0.45]{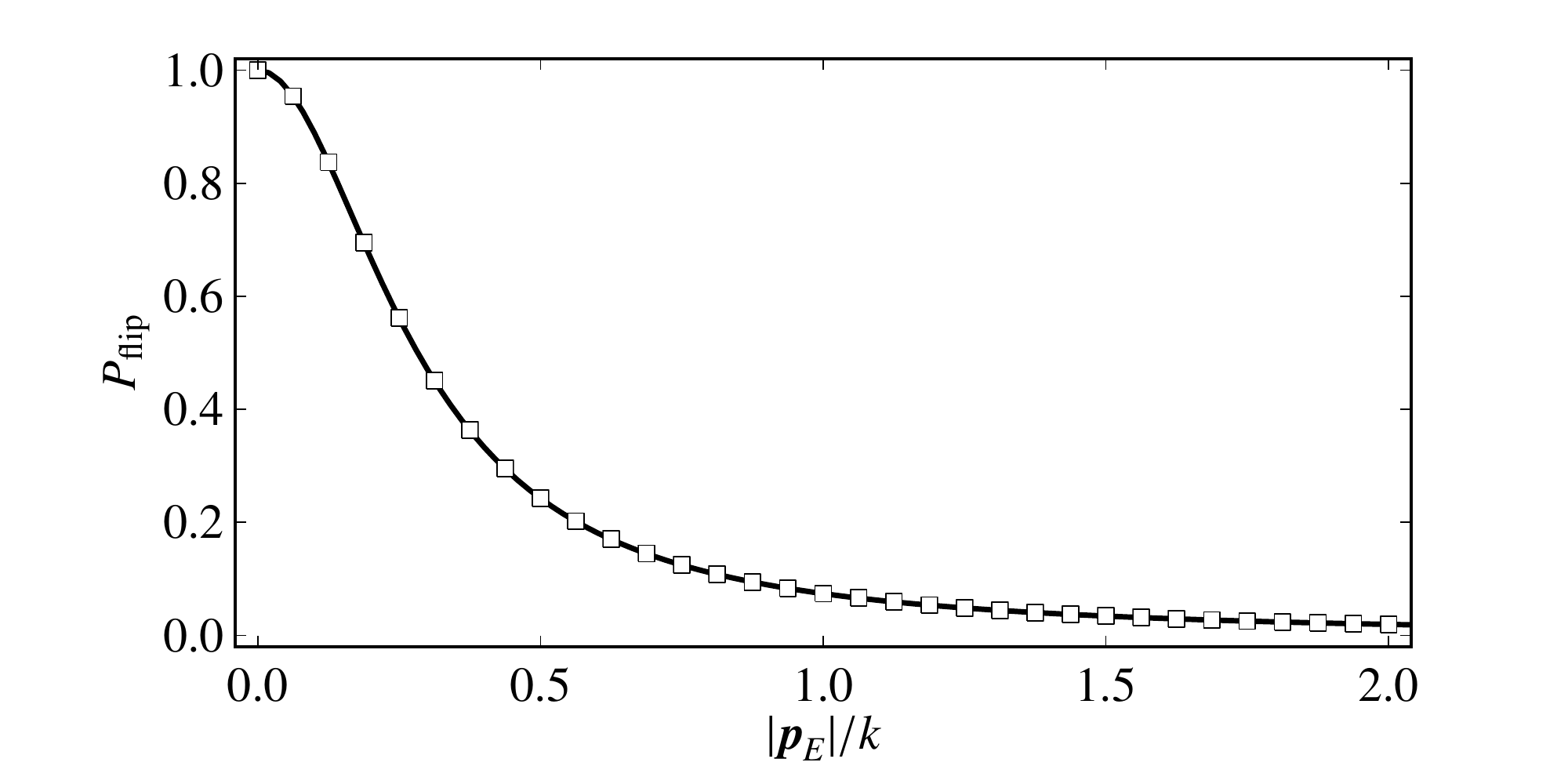}
  \caption{The spin-flip probability $P_\text{flip}$ as a function of
    the electron momentum $|\vec{p}_E|$ in electric field direction. The
    solid black line is given by (\ref{eq:spin-flip-probability}). White
    squares result from simulations with the Dirac equation
    (\ref{eq:FSME-dirac-equation}). All simulation parameters except
    $\vec{p}_E$ are the same as those for Fig.~\ref{fig:spin_flip}.}
  \label{fig:alpha_0_3}
\end{figure}

Figure~\ref{fig:alpha_0_3} compares the spin-flip probability as a
function of $|\vec p_E|/k$, as obtained from the numerical solution of
the Dirac equation~(\ref{eq:FSME-dirac-equation}), with our analytical
result (\ref{eq:spin-flip-probability}).  The initial electron momentum
in laser propagation direction $\vec p_k$ is adjusted according to
equation (\ref{eq:gen_Bragg}) for each value of $\vec p_E$. The
analytical formula (\ref{eq:spin-flip-probability}) shows very good
agreement with the numerical data.

\paragraph{Considerations on nonrelativistic theory}

The spin-flip probability (\ref{eq:spin-flip-probability}) can be
understood on a qualitative level by analyzing the leading order of the
Foldy-Wouthuysen expansion \cite{Foldy:Wouthuysen:1950:nr_limit} of the
Dirac equation (\ref{eq:Dirac}) which equals the nonrelativistic Pauli
equation.  This equation features two coupling terms which are linear in
the fields, namely ${e}\vec A\cdot \vec p/({mc})$ and
${e}\vec\sigma\cdot\vec B/({2mc})$, where $\vec\sigma$ denotes the
vector of Pauli matrices. Both terms give rise to couplings between
adjacent electron momentum components (differing by $\pm\vec k$), with
the first term preserving and the second term flipping the electron
spin. Note that such nearest-neighbor couplings are necessarily involved
in Kapitza-Dirac scattering with an odd number of photons.  The relative
strength of the $\vec A\cdot\vec p$ term as compared with the
$\vec\sigma\cdot\vec B$ term is just $2 |\vec p_E| / k$. This results in
the spin-flip probability
\begin{equation}
  \label{eq:spin-flip-probability_nr}
  P_\text{flip, nonrel.} =
  \frac{1}{
    4\left(
      \frac{|\vec p_E|}{k}
    \right)^2 + 1} \,,
\end{equation}
which agrees with (\ref{eq:spin-flip-probability}) up to a scale
parameter $25/8$. The probability (\ref{eq:spin-flip-probability_nr})
can also be derived more rigorously via time-dependent perturbation
theory for the Pauli equation, which yields the nonrelativistic Rabi
frequency
\begin{equation}
  \label{eq:Rabi_PE}
  \Omega_{R\text{, nonrel.}} = \frac{243}{128}\Omega_0
  \sqrt{4\left(
      \frac{|\vec p_E|}{k}
    \right)^2 + 1}\,.
\end{equation}
Note that the relativistic and nonrelativistic spin-flip probabilities
(\ref{eq:spin-flip-probability}) and (\ref{eq:spin-flip-probability_nr})
and the relativistic and nonrelativistic Rabi frequencies
(\ref{eq:Rabi_DE}) and (\ref{eq:Rabi_PE}) agree qualitatively. However,
the nonrelativistic expressions cannot be recovered from the
relativistic ones by taking a nonrelativistic limit.  This is a
consequence of the three-photon Bragg condition (\ref{eq:gen_Bragg}) that
enforces a relativistic photon momentum and/or a relativistic electron
momentum highlighting the relativistic nature of the three-photon
Kapitza-Dirac effect.

\paragraph{The role of the spin}

Equation (\ref{eq:c_3_2}) indicates that spin-preserving transitions in
the \mbox{three-photon} Kapitza-Dirac effect are completely suppressed
for setups with $\vec p\perp\vec E$. This means that under such
conditions the scattering is rendered possible only because the electron
does carry spin, which clearly demonstrates the pivotal role the
electron spin can play in Kapitza-Dirac scattering processes.

\begin{figure}
  \centering
  \hspace{\fill}%
  \includegraphics[scale=0.45]{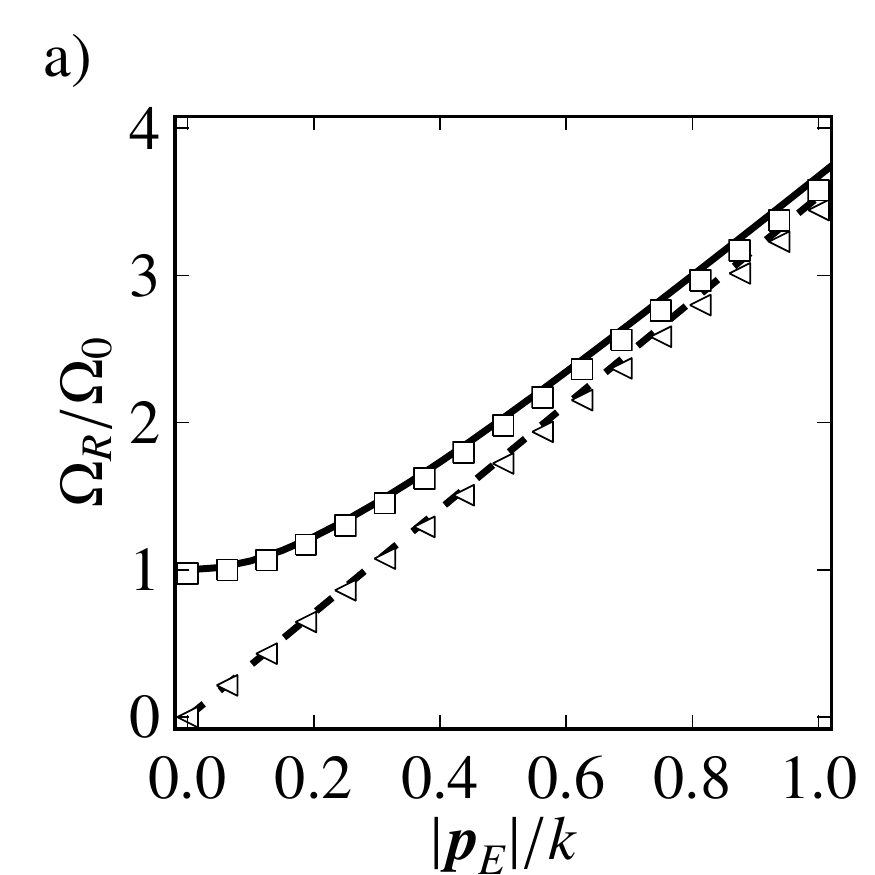}%
  \hspace{\fill}\hspace{\fill}%
  \includegraphics[scale=0.45]{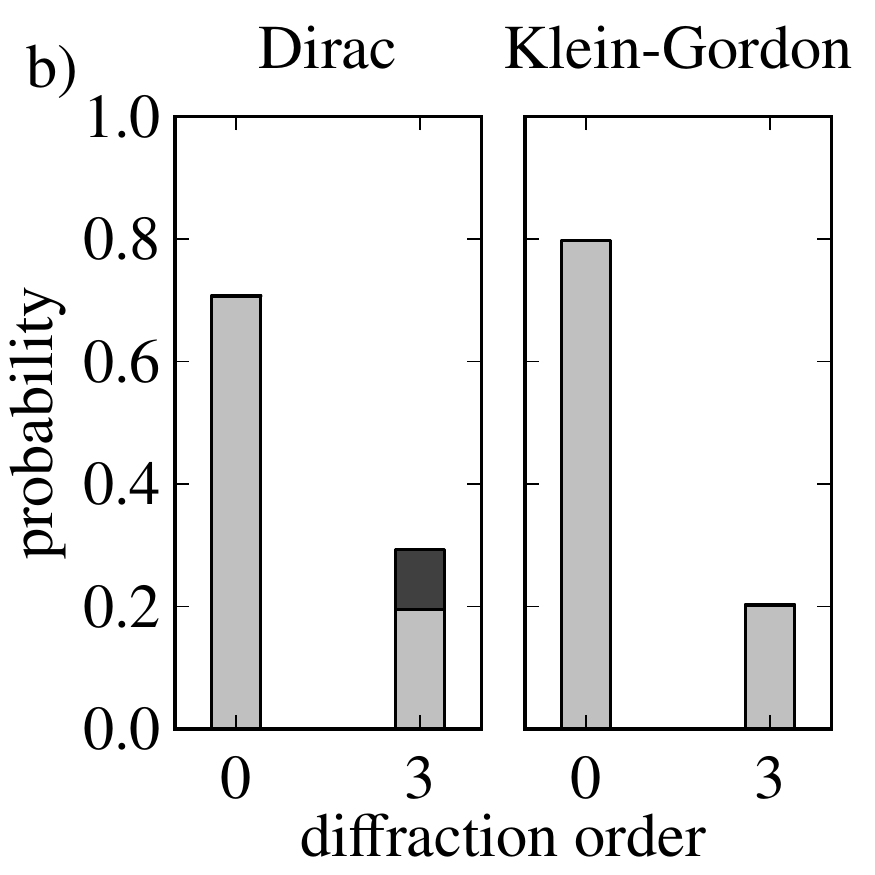}
  \hspace{\fill}\mbox{}%
  \caption{Panel~(a): The Rabi frequency $\Omega_R$ as a function of the
    electron momentum $|\vec{p}_E|$ in electric field direction for the
    Dirac equation (squares) and the Klein-Gordon equation
    (triangles). The solid black line is given by (\ref{eq:Rabi_DE}) and
    the dashed black line is given by (\ref{eq:Rabi_KGE}). Panel~(b):
    The diffraction probability after an interaction time of 0.36\,fs
    for particles with and without spin for parameters as in
    Fig.~\ref{fig:spin_flip}.  The light (dark) gray bars represent the
    spin-up (spin-down) probabilities. In the case of the Klein-Gordon
    equation there is no spin degree of freedom and, therefore, no dark
    gray bars appear. Note that the diffraction probability depends on
    the spin degree of freedom.}
  \label{fig:Omega_R}
\end{figure}

The above argument suggests that for a spinless particle with $\vec
p\perp \vec E$ the \mbox{three-photon} channel of Kapitza-Dirac
scattering is not accessible at all.  This expectation is confirmed by
Fig.~\ref{fig:Omega_R}\,(a). It compares numerical results on the Rabi
frequency as following from the Dirac equation
(\ref{eq:FSME-dirac-equation}) with corresponding numbers that we
obtained by solving the time-dependent Klein-Gordon equation
\footnote{An expansion of the wave function into momentum eigenfunctions
  of the Klein-Gordon equation in Hamiltonian form
  \cite{Feshbach:Villars:1954:Relativistic_Wave_Mechanics,*Ruf:2009:Klein-Gordon_Splitoperator}
  yields a system of ordinary differential equations similar to
  (\ref{eq:FSME-dirac-equation}).}.
\vphantom{\cite{Feshbach:Villars:1954:Relativistic_Wave_Mechanics,*Ruf:2009:Klein-Gordon_Splitoperator}}%
The predictions significantly differ in the limit $\vec p_E \to
0$. While the Rabi frequency converges to a finite value for spin-half
particles [see also (\ref{eq:Rabi_DE})], it approaches zero for spinless
particles, indicating that the scattering channel closes indeed. In the
spinless case, the spin-flipped channel (\ref{eq:c_3_2_down}) is missing,
which consequently yields the Rabi frequency
\begin{equation}
  \label{eq:Rabi_KGE}
  \Omega_{R,\,\text{spinless}} = 
  \Omega_0\frac{5}{\sqrt{2}}\frac{|\vec p_E|}{k}\,.
\end{equation}
For nonzero values of $\vec p_E$, also Klein-Gordon particles may be
scattered. But the scattering probability still may be considerably
different from the Dirac case, as the example in
Fig.~\ref{fig:Omega_R}\,(b) shows.

\enlargethispage{3ex}

\paragraph{Experimental realization}

An experimental realization of the \mbox{three-photon} Kapitza-Dirac
effect may utilize intense photon beams at near-future \mbox{X-ray}
laser facilities to form standing waves. In our numerical simulations,
we assumed 3.1\,keV photons as envisaged, for example, at the European
\mbox{X-ray} free electron laser facility (XFEL)
\cite{Altarell:etal:2007:XFEL}, which is currently under
construction. The design value of the peak power at this photon energy
is 80\,GW.  Assuming a focus diameter of 7\,nm
\cite{Mimura:2010:10nm-x-ray-focusing}, a field intensity of about $2
\times 10^{23}\,\text{W}/\text{cm}^2$ results. Laser pulses with
duration of about half a Rabi period [which is about 1\,fs for the
parameters in Fig.~\ref{fig:spin_flip})] are required for experimental
realization \cite{Zholents:Penn:2005:attosecond_x-ray-pulses}.  Since
the Rabi frequency is much lower than the laser frequency, the photon
energy and the electron momentum must be fine-tuned to achieve a
resonant transition. Numerical simulations indicate that only electrons
whose momentum varies by $0.1\,\textrm{keV}/c$ around the mean value of
$176\,\textrm{keV}/c$ are diffracted. The photon pulse of the European
XFEL with a seeded beam of a primary undulator is expected to be
coherent, featuring a photon energy uncertainty far below 0.1\,keV
\cite{Geloni:Kocharyan:Saldin:2011:self-seeding_X-ray_FELs}. We note
that the electron may lose energy due to spontaneous photoemission with
resulting quantitative modification of the presented results.
Spontaneous emission (scaling with $|\vec E|^2$), however, is
substantially suppressed as compared with the very fast momentum
transfer through the three-photon Kapitza-Dirac effect which takes place
on a femtosecond time scale during a Rabi period ($1/\Omega_R \sim
1/|\vec E|^3$). A numerical solution of the Landau-Lifshitz equation
\cite{Landau:Lifshitz:1975:2,*Keitel:etal:1998:Radiative_reaction,*DiPiazza:etal:2009:Radiation_Reaction}
indicates that the momentum transfer into laser propagation direction
caused by spontaneous emission is sufficiently small in order not to
violate the resonance condition (\ref{eq:gen_Bragg}) for the current
parameters.  Finally, the electron beam is diffracted almost in the
electron propagation direction by $3 \times 3.1\,\text{keV}/c$.
Therefore, a spectrometer with a resolution below $10\,\text{keV}/c$
should be able to separate the diffracted electron beam from the not
diffracted one.  In the diffracted beam, about one out of three electrons
are spin flipped in the case of the scenario in
Fig.~\ref{fig:spin_flip}. This spin-flip fraction is independent of the
interaction time $T$ of the electron with the laser, in accordance with
(\ref{eq:spin-flip-probability}).

\paragraph{Conclusions}

Pronounced spin effects in Kapitza-Dirac scattering involving three
\mbox{X-ray} laser photons interacting with a weakly relativistic
electron beam have been revealed.  To this end, we deduced a generalized
Bragg condition and developed a theoretical description of the quantum
dynamics based on the Dirac equation. The process features
characteristic Rabi oscillations and a competition between
spin-preserving and spin-flipping nearest-neighbor couplings. The
spin-flipping transition becomes dominant in the limit of small angles
of inclination, where \mbox{three-photon} Kapitza-Dirac scattering
crucially relies on the nonzero electron spin. Our predictions may be
tested with the aid of near-future high-intensity XFEL sources.

\bibliography{spin_kde}

\begin{thebibliography}{47}%
\makeatletter
\providecommand \@ifxundefined [1]{%
 \@ifx{#1\undefined}
}%
\providecommand \@ifnum [1]{%
 \ifnum #1\expandafter \@firstoftwo
 \else \expandafter \@secondoftwo
 \fi
}%
\providecommand \@ifx [1]{%
 \ifx #1\expandafter \@firstoftwo
 \else \expandafter \@secondoftwo
 \fi
}%
\providecommand \natexlab [1]{#1}%
\providecommand \enquote  [1]{``#1''}%
\providecommand \bibnamefont  [1]{#1}%
\providecommand \bibfnamefont [1]{#1}%
\providecommand \citenamefont [1]{#1}%
\providecommand \href@noop [0]{\@secondoftwo}%
\providecommand \href [0]{\begingroup \@sanitize@url \@href}%
\providecommand \@href[1]{\@@startlink{#1}\@@href}%
\providecommand \@@href[1]{\endgroup#1\@@endlink}%
\providecommand \@sanitize@url [0]{\catcode `\\12\catcode `\$12\catcode
  `\&12\catcode `\#12\catcode `\^12\catcode `\_12\catcode `\%12\relax}%
\providecommand \@@startlink[1]{}%
\providecommand \@@endlink[0]{}%
\providecommand \url  [0]{\begingroup\@sanitize@url \@url }%
\providecommand \@url [1]{\endgroup\@href {#1}{\urlprefix }}%
\providecommand \urlprefix  [0]{URL }%
\providecommand \Eprint [0]{\href }%
\providecommand \doibase [0]{http://dx.doi.org/}%
\providecommand \selectlanguage [0]{\@gobble}%
\providecommand \bibinfo  [0]{\@secondoftwo}%
\providecommand \bibfield  [0]{\@secondoftwo}%
\providecommand \translation [1]{[#1]}%
\providecommand \BibitemOpen [0]{}%
\providecommand \bibitemStop [0]{}%
\providecommand \bibitemNoStop [0]{.\EOS\space}%
\providecommand \EOS [0]{\spacefactor3000\relax}%
\providecommand \BibitemShut  [1]{\csname bibitem#1\endcsname}%
\let\auto@bib@innerbib\@empty
\bibitem [{\citenamefont {Kapitza}\ and\ \citenamefont
  {Dirac}(1933)}]{Kapitza:Dirac:1933:effect}%
  \BibitemOpen
  \bibfield  {author} {\bibinfo {author} {\bibfnamefont {P.~L.}\ \bibnamefont
  {Kapitza}}\ \bibnamefont {and}\ \bibinfo {author} {\bibfnamefont {P.~A.~M.}\
  \bibnamefont {Dirac}},\ }\href {\doibase 10.1017/S0305004100011105}
  {\bibfield  {journal} {\bibinfo  {journal} {Math. Proc. Cambridge Philos.
  Soc.}\ }\textbf {\bibinfo {volume} {29}},\ \bibinfo {pages} {297} (\bibinfo
  {year} {1933})}\BibitemShut {NoStop}%
\bibitem [{\citenamefont
  {Batelaan}(2007)}]{Batelaan:2007:Illuminating_Kapitza-Dirac}%
  \BibitemOpen
  \bibfield  {author} {\bibinfo {author} {\bibfnamefont {H.}~\bibnamefont
  {Batelaan}},\ }\href {\doibase 10.1103/RevModPhys.79.929} {\bibfield
  {journal} {\bibinfo  {journal} {Rev. Mod. Phys.}\ }\textbf {\bibinfo {volume}
  {79}},\ \bibinfo {pages} {929} (\bibinfo {year} {2007})}\BibitemShut
  {NoStop}%
\bibitem [{\citenamefont {Freimund}\ \emph {et~al.}(2001)\citenamefont
  {Freimund}, \citenamefont {Aflatooni},\ and\ \citenamefont
  {Batelaan}}]{Freimund:etal:2001:Observation_Kapitza-Dirac}%
  \BibitemOpen
  \bibfield  {author} {\bibinfo {author} {\bibfnamefont {D.~L.}\ \bibnamefont
  {Freimund}}, \bibinfo {author} {\bibfnamefont {K.}~\bibnamefont
  {Aflatooni}},\ \bibnamefont {and}\ \bibinfo {author} {\bibfnamefont
  {H.}~\bibnamefont {Batelaan}},\ }\href {\doibase 10.1038/35093065} {\bibfield
   {journal} {\bibinfo  {journal} {Nature}\ }\textbf {\bibinfo {volume}
  {413}},\ \bibinfo {pages} {142} (\bibinfo {year} {2001})}\BibitemShut
  {NoStop}%
\bibitem [{\citenamefont {Freimund}\ and\ \citenamefont
  {Batelaan}(2002)}]{Freimund:Batelaan:2002:Bragg}%
  \BibitemOpen
  \bibfield  {author} {\bibinfo {author} {\bibfnamefont {D.~L.}\ \bibnamefont
  {Freimund}}\ \bibnamefont {and}\ \bibinfo {author} {\bibfnamefont
  {H.}~\bibnamefont {Batelaan}},\ }\href {\doibase
  10.1103/PhysRevLett.89.283602} {\bibfield  {journal} {\bibinfo  {journal}
  {Phys. Rev. Lett.}\ }\textbf {\bibinfo {volume} {89}},\ \bibinfo {pages}
  {283602} (\bibinfo {year} {2002})}\BibitemShut {NoStop}%
\bibitem [{\citenamefont {Smirnova}\ \emph {et~al.}(2004)\citenamefont
  {Smirnova}, \citenamefont {Freimund}, \citenamefont {Batelaan},\ and\
  \citenamefont {Ivanov}}]{Smirnova:etal:2004:two_color_Kapitza-Dirac}%
  \BibitemOpen
  \bibfield  {author} {\bibinfo {author} {\bibfnamefont {O.}~\bibnamefont
  {Smirnova}}, \bibinfo {author} {\bibfnamefont {D.~L.}\ \bibnamefont
  {Freimund}}, \bibinfo {author} {\bibfnamefont {H.}~\bibnamefont {Batelaan}},\
  \bibnamefont {and}\ \bibinfo {author} {\bibfnamefont {M.}~\bibnamefont
  {Ivanov}},\ }\href {\doibase 10.1103/PhysRevLett.92.223601} {\bibfield
  {journal} {\bibinfo  {journal} {Phys. Rev. Lett.}\ }\textbf {\bibinfo
  {volume} {92}},\ \bibinfo {pages} {223601} (\bibinfo {year}
  {2004})}\BibitemShut {NoStop}%
\bibitem [{\citenamefont {Li}\ \emph {et~al.}(2004)\citenamefont {Li},
  \citenamefont {Zhang}, \citenamefont {Xu}, \citenamefont {Fu}, \citenamefont
  {Guo},\ and\ \citenamefont {Freeman}}]{Li:etal:2004:Theory_Kapitza-Dirac}%
  \BibitemOpen
  \bibfield  {author} {\bibinfo {author} {\bibfnamefont {X.}~\bibnamefont
  {Li}}, \bibinfo {author} {\bibfnamefont {J.}~\bibnamefont {Zhang}}, \bibinfo
  {author} {\bibfnamefont {Z.}~\bibnamefont {Xu}}, \bibinfo {author}
  {\bibfnamefont {P.}~\bibnamefont {Fu}}, \bibinfo {author} {\bibfnamefont
  {D.-S.}\ \bibnamefont {Guo}},\ \bibnamefont {and}\ \bibinfo {author}
  {\bibfnamefont {R.~R.}\ \bibnamefont {Freeman}},\ }\href {\doibase
  10.1103/PhysRevLett.92.233603} {\bibfield  {journal} {\bibinfo  {journal}
  {Phys. Rev. Lett.}\ }\textbf {\bibinfo {volume} {92}},\ \bibinfo {pages}
  {233603} (\bibinfo {year} {2004})}\BibitemShut {NoStop}%
\bibitem [{\citenamefont
  {Sancho}(2010)}]{Sancho:2010:two-particle_Kapitza-Dirac}%
  \BibitemOpen
  \bibfield  {author} {\bibinfo {author} {\bibfnamefont {P.}~\bibnamefont
  {Sancho}},\ }\href {\doibase 10.1103/PhysRevA.82.033814} {\bibfield
  {journal} {\bibinfo  {journal} {Phys. Rev. A}\ }\textbf {\bibinfo {volume}
  {82}},\ \bibinfo {pages} {033814} (\bibinfo {year} {2010})}\BibitemShut
  {NoStop}%
\bibitem [{\citenamefont {Fedorov}(1967)}]{Fedorov:1967:Kapitza_Dirac}%
  \BibitemOpen
  \bibfield  {author} {\bibinfo {author} {\bibfnamefont {M.~V.}\ \bibnamefont
  {Fedorov}},\ }\href@noop {} {\bibfield  {journal} {\bibinfo  {journal} {Sov.
  Phys.--JETP}\ }\textbf {\bibinfo {volume} {25}},\ \bibinfo {pages} {952}
  (\bibinfo {year} {1967})}\BibitemShut {NoStop}%
\bibitem [{\citenamefont {Gush}\ and\ \citenamefont
  {Gush}(1971)}]{Gush:1971:Electron_Scattering}%
  \BibitemOpen
  \bibfield  {author} {\bibinfo {author} {\bibfnamefont {R.}~\bibnamefont
  {Gush}}\ \bibnamefont {and}\ \bibinfo {author} {\bibfnamefont {H.~P.}\
  \bibnamefont {Gush}},\ }\href {\doibase 10.1103/PhysRevD.3.1712} {\bibfield
  {journal} {\bibinfo  {journal} {Phys. Rev. D}\ }\textbf {\bibinfo {volume}
  {3}},\ \bibinfo {pages} {1712} (\bibinfo {year} {1971})}\BibitemShut
  {NoStop}%
\bibitem [{\citenamefont {Fedorov}(1974)}]{Fedorov:1974:Stimulated_scattering}%
  \BibitemOpen
  \bibfield  {author} {\bibinfo {author} {\bibfnamefont {M.~V.}\ \bibnamefont
  {Fedorov}},\ }\href {\doibase 10.1016/0030-4018(74)90392-7} {\bibfield
  {journal} {\bibinfo  {journal} {Opt. Commun.}\ }\textbf {\bibinfo {volume}
  {12}},\ \bibinfo {pages} {205} (\bibinfo {year} {1974})}\BibitemShut
  {NoStop}%
\bibitem [{\citenamefont {Bucksbaum}\ \emph {et~al.}(1988)\citenamefont
  {Bucksbaum}, \citenamefont {Schumacher},\ and\ \citenamefont
  {Bashkansky}}]{Bucksbaum:etal:1988:Kapitza-Dirac_Effect}%
  \BibitemOpen
  \bibfield  {author} {\bibinfo {author} {\bibfnamefont {P.~H.}\ \bibnamefont
  {Bucksbaum}}, \bibinfo {author} {\bibfnamefont {D.~W.}\ \bibnamefont
  {Schumacher}},\ \bibnamefont {and}\ \bibinfo {author} {\bibfnamefont
  {M.}~\bibnamefont {Bashkansky}},\ }\href {\doibase
  10.1103/PhysRevLett.61.1182} {\bibfield  {journal} {\bibinfo  {journal}
  {Phys. Rev. Lett.}\ }\textbf {\bibinfo {volume} {61}},\ \bibinfo {pages}
  {1182} (\bibinfo {year} {1988})}\BibitemShut {NoStop}%
\bibitem [{\citenamefont {Gould}\ \emph {et~al.}(1986)\citenamefont {Gould},
  \citenamefont {Ruff},\ and\ \citenamefont
  {Pritchard}}]{Gould:etal:1986:Diffraction_of_atoms}%
  \BibitemOpen
  \bibfield  {author} {\bibinfo {author} {\bibfnamefont {P.~L.}\ \bibnamefont
  {Gould}}, \bibinfo {author} {\bibfnamefont {G.~A.}\ \bibnamefont {Ruff}},\
  \bibnamefont {and}\ \bibinfo {author} {\bibfnamefont {D.~E.}\ \bibnamefont
  {Pritchard}},\ }\href {\doibase 10.1103/PhysRevLett.56.827} {\bibfield
  {journal} {\bibinfo  {journal} {Phys. Rev. Lett.}\ }\textbf {\bibinfo
  {volume} {56}},\ \bibinfo {pages} {827} (\bibinfo {year} {1986})}\BibitemShut
  {NoStop}%
\bibitem [{\citenamefont {Adams}\ \emph {et~al.}(1994)\citenamefont {Adams},
  \citenamefont {Sigel},\ and\ \citenamefont
  {Mlynek}}]{Adams:etal:1994:Atom_optics}%
  \BibitemOpen
  \bibfield  {author} {\bibinfo {author} {\bibfnamefont {C.}~\bibnamefont
  {Adams}}, \bibinfo {author} {\bibfnamefont {M.}~\bibnamefont {Sigel}},\
  \bibnamefont {and}\ \bibinfo {author} {\bibfnamefont {J.}~\bibnamefont
  {Mlynek}},\ }\href {\doibase 10.1016/0370-1573(94)90066-3} {\bibfield
  {journal} {\bibinfo  {journal} {Phys. Rep.}\ }\textbf {\bibinfo {volume}
  {240}},\ \bibinfo {pages} {143} (\bibinfo {year} {1994})}\BibitemShut
  {NoStop}%
\bibitem [{\citenamefont {Freyberger}\ \emph {et~al.}(1999)\citenamefont
  {Freyberger}, \citenamefont {Herkommer}, \citenamefont {Kr\"ahmer},
  \citenamefont {Mayr},\ and\ \citenamefont
  {Schleich}}]{Freyberger:etal:1999:Atom_Optics}%
  \BibitemOpen
  \bibfield  {author} {\bibinfo {author} {\bibfnamefont {M.}~\bibnamefont
  {Freyberger}}, \bibinfo {author} {\bibfnamefont {A.}~\bibnamefont
  {Herkommer}}, \bibinfo {author} {\bibfnamefont {D.}~\bibnamefont
  {Kr\"ahmer}}, \bibinfo {author} {\bibfnamefont {E.}~\bibnamefont {Mayr}},\
  \bibnamefont {and}\ \bibinfo {author} {\bibfnamefont {W.}~\bibnamefont
  {Schleich}},\ }in\ \href {\doibase 10.1016/S1049-250X(08)60220-7} {\emph
  {\bibinfo {booktitle} {Adv. At., Mol., Opt. Phys.}}},\ Vol.~\bibinfo {volume}
  {41},\ \bibinfo {editor} {edited by\ \bibinfo {editor} {\bibfnamefont
  {B.}~\bibnamefont {Bederson}}\ \bibnamefont {and}\ \bibinfo {editor}
  {\bibfnamefont {H.}~\bibnamefont {Walther}}}\ (\bibinfo  {publisher}
  {Academic Press},\ \bibinfo {address} {New York},\ \bibinfo {year} {1999})\
  pp.\ \bibinfo {pages} {143--180}\BibitemShut {NoStop}%
\bibitem [{\citenamefont {Altarelli}\ \emph {et~al.}(2007)\citenamefont
  {Altarelli}, \citenamefont {Brinkmann}, \citenamefont {Chergui},
  \citenamefont {Decking}, \citenamefont {Dobson}, \citenamefont
  {D{\"u}sterer}, \citenamefont {Gr{\"u}bel}, \citenamefont {Graeff},
  \citenamefont {Graafsma}, \citenamefont {Hajdu}, \citenamefont {Marangos},
  \citenamefont {Pfl{\"u}ger}, \citenamefont {Redlin}, \citenamefont {Riley},
  \citenamefont {Robinson}, \citenamefont {Rossbach}, \citenamefont {Schwarz},
  \citenamefont {Tiedtke}, \citenamefont {Tschentscher}, \citenamefont
  {Vartaniants}, \citenamefont {Wabnitz}, \citenamefont {Weise}, \citenamefont
  {Wichmann}, \citenamefont {Witte}, \citenamefont {Wolf}, \citenamefont
  {Wulff},\ and\ \citenamefont {Yurkov}}]{Altarell:etal:2007:XFEL}%
  \BibitemOpen
  \bibinfo {editor} {\bibfnamefont {M.}~\bibnamefont {Altarelli}}, \bibinfo
  {editor} {\bibfnamefont {R.}~\bibnamefont {Brinkmann}}, \bibinfo {editor}
  {\bibfnamefont {M.}~\bibnamefont {Chergui}}, \bibinfo {editor} {\bibfnamefont
  {W.}~\bibnamefont {Decking}}, \bibinfo {editor} {\bibfnamefont
  {B.}~\bibnamefont {Dobson}}, \bibinfo {editor} {\bibfnamefont
  {S.}~\bibnamefont {D{\"u}sterer}}, \bibinfo {editor} {\bibfnamefont
  {G.}~\bibnamefont {Gr{\"u}bel}}, \bibinfo {editor} {\bibfnamefont
  {W.}~\bibnamefont {Graeff}}, \bibinfo {editor} {\bibfnamefont
  {H.}~\bibnamefont {Graafsma}}, \bibinfo {editor} {\bibfnamefont
  {J.}~\bibnamefont {Hajdu}} \emph {et~al.},\ eds.,\ \href@noop {} {\emph
  {\bibinfo {title} {The European X-Ray Free-Electron Laser Technical design
  report}}}\ (\bibinfo  {publisher} {DESY XFEL Project Group European XFEL
  Project Team Deutsches Elektronen-Synchrotron Member of the Helmholtz
  Association},\ \bibinfo {address} {Hamburg},\ \bibinfo {year}
  {2007})\BibitemShut {NoStop}%
\bibitem [{\citenamefont {McNeil}\ and\ \citenamefont
  {Thompson}(2010)}]{McNeil:Thompson:2010:XFEL}%
  \BibitemOpen
  \bibfield  {author} {\bibinfo {author} {\bibfnamefont {B.~W.~J.}\
  \bibnamefont {McNeil}}\ \bibnamefont {and}\ \bibinfo {author} {\bibfnamefont
  {N.~R.}\ \bibnamefont {Thompson}},\ }\href {\doibase
  10.1038/nphoton.2010.239} {\bibfield  {journal} {\bibinfo  {journal} {Nature
  Photon.}\ }\textbf {\bibinfo {volume} {4}},\ \bibinfo {pages} {814} (\bibinfo
  {year} {2010})}\BibitemShut {NoStop}%
\bibitem [{\citenamefont {Emma}\ \emph {et~al.}(2010)\citenamefont {Emma},
  \citenamefont {Akre}, \citenamefont {Arthur}, \citenamefont {Bionta},
  \citenamefont {Bostedt}, \citenamefont {Bozek}, \citenamefont {Brachmann},
  \citenamefont {Bucksbaum}, \citenamefont {Coffee}, \citenamefont {Decker},
  \citenamefont {Ding}, \citenamefont {Dowell}, \citenamefont {Edstrom},
  \citenamefont {Fisher}, \citenamefont {Frisch}, \citenamefont {Gilevich},
  \citenamefont {Hastings}, \citenamefont {Hays}, \citenamefont {Hering},
  \citenamefont {Huang}, \citenamefont {Iverson}, \citenamefont {Loos},
  \citenamefont {Messerschmidt}, \citenamefont {Miahnahri}, \citenamefont
  {Moeller}, \citenamefont {Nuhn}, \citenamefont {Pile}, \citenamefont
  {Ratner}, \citenamefont {Rzepiela}, \citenamefont {Schultz}, \citenamefont
  {Smith}, \citenamefont {Stefan}, \citenamefont {Tompkins}, \citenamefont
  {Turner}, \citenamefont {Welch}, \citenamefont {White}, \citenamefont {Wu},
  \citenamefont {Yocky},\ and\ \citenamefont
  {Galayda}}]{Emma:etal:2010:First_lasing}%
  \BibitemOpen
  \bibfield  {author} {\bibinfo {author} {\bibfnamefont {P.}~\bibnamefont
  {Emma}}, \bibinfo {author} {\bibfnamefont {R.}~\bibnamefont {Akre}}, \bibinfo
  {author} {\bibfnamefont {J.}~\bibnamefont {Arthur}}, \bibinfo {author}
  {\bibfnamefont {R.}~\bibnamefont {Bionta}}, \bibinfo {author} {\bibfnamefont
  {C.}~\bibnamefont {Bostedt}}, \bibinfo {author} {\bibfnamefont
  {J.}~\bibnamefont {Bozek}}, \bibinfo {author} {\bibfnamefont
  {A.}~\bibnamefont {Brachmann}}, \bibinfo {author} {\bibfnamefont
  {P.}~\bibnamefont {Bucksbaum}}, \bibinfo {author} {\bibfnamefont
  {R.}~\bibnamefont {Coffee}}, \bibinfo {author} {\bibfnamefont {F.-J.}\
  \bibnamefont {Decker}} \emph {et~al.},\ }\href {\doibase
  10.1038/nphoton.2010.176} {\bibfield  {journal} {\bibinfo  {journal} {Nature
  Photon.}\ }\textbf {\bibinfo {volume} {4}},\ \bibinfo {pages} {641} (\bibinfo
  {year} {2010})}\BibitemShut {NoStop}%
\bibitem [{\citenamefont {Mourou}\ and\ \citenamefont
  {Tajima}(2011)}]{Mourou:Tajima:2011:More_intense}%
  \BibitemOpen
  \bibfield  {author} {\bibinfo {author} {\bibfnamefont {G.}~\bibnamefont
  {Mourou}}\ \bibnamefont {and}\ \bibinfo {author} {\bibfnamefont
  {T.}~\bibnamefont {Tajima}},\ }\href {\doibase 10.1126/science.1200292}
  {\bibfield  {journal} {\bibinfo  {journal} {Science}\ }\textbf {\bibinfo
  {volume} {331}},\ \bibinfo {pages} {41} (\bibinfo {year} {2011})}\BibitemShut
  {NoStop}%
\bibitem [{\citenamefont {Haroutunian}\ and\ \citenamefont
  {Avetissian}(1975)}]{Haroutunian:Avetissian:1975:analogue_KGE}%
  \BibitemOpen
  \bibfield  {author} {\bibinfo {author} {\bibfnamefont {V.~M.}\ \bibnamefont
  {Haroutunian}}\ \bibnamefont {and}\ \bibinfo {author} {\bibfnamefont {H.~K.}\
  \bibnamefont {Avetissian}},\ }\href {\doibase 10.1016/0375-9601(75)90628-3}
  {\bibfield  {journal} {\bibinfo  {journal} {Phys. Lett. A}\ }\textbf
  {\bibinfo {volume} {51}},\ \bibinfo {pages} {320} (\bibinfo {year}
  {1975})}\BibitemShut {NoStop}%
\bibitem [{\citenamefont {Federov}\ and\ \citenamefont
  {McIver}(1980)}]{Federov:McIver:1980:compton_scattering}%
  \BibitemOpen
  \bibfield  {author} {\bibinfo {author} {\bibfnamefont {M.~V.}\ \bibnamefont
  {Federov}}\ \bibnamefont {and}\ \bibinfo {author} {\bibfnamefont {J.~K.}\
  \bibnamefont {McIver}},\ }\href {\doibase 10.1016/0030-4018(80)90341-7}
  {\bibfield  {journal} {\bibinfo  {journal} {Opt. Commun.}\ }\textbf {\bibinfo
  {volume} {32}},\ \bibinfo {pages} {179} (\bibinfo {year} {1980})}\BibitemShut
  {NoStop}%
\bibitem [{\citenamefont {Rosenberg}(2004)}]{Rosenberg:2004:Extended_theory}%
  \BibitemOpen
  \bibfield  {author} {\bibinfo {author} {\bibfnamefont {L.}~\bibnamefont
  {Rosenberg}},\ }\href {\doibase 10.1103/PhysRevA.70.023401} {\bibfield
  {journal} {\bibinfo  {journal} {Phys. Rev. A}\ }\textbf {\bibinfo {volume}
  {70}},\ \bibinfo {pages} {023401} (\bibinfo {year} {2004})}\BibitemShut
  {NoStop}%
\bibitem [{\citenamefont {Freimund}\ and\ \citenamefont
  {Batelaan}(2003)}]{Freimund:Batelaan:2003:Stern-Gerlach}%
  \BibitemOpen
  \bibfield  {author} {\bibinfo {author} {\bibfnamefont {D.~L.}\ \bibnamefont
  {Freimund}}\ \bibnamefont {and}\ \bibinfo {author} {\bibfnamefont
  {H.}~\bibnamefont {Batelaan}},\ }\href@noop {} {\bibfield  {journal}
  {\bibinfo  {journal} {Laser Phys.}\ }\textbf {\bibinfo {volume} {13}},\
  \bibinfo {pages} {892} (\bibinfo {year} {2003})}\BibitemShut {NoStop}%
\bibitem [{\citenamefont {Krekora}\ \emph {et~al.}(2001)\citenamefont
  {Krekora}, \citenamefont {Su},\ and\ \citenamefont
  {Grobe}}]{Krekora:Su:Grobe:2001:spin_of_relativistic_electrons}%
  \BibitemOpen
  \bibfield  {author} {\bibinfo {author} {\bibfnamefont {P.}~\bibnamefont
  {Krekora}}, \bibinfo {author} {\bibfnamefont {Q.}~\bibnamefont {Su}},\
  \bibnamefont {and}\ \bibinfo {author} {\bibfnamefont {R.}~\bibnamefont
  {Grobe}},\ }\href {\doibase 10.1088/0953-4075/34/14/303} {\bibfield
  {journal} {\bibinfo  {journal} {J. Phys. B: At., Mol. Opt. Phys.}\ }\textbf
  {\bibinfo {volume} {34}},\ \bibinfo {pages} {2795} (\bibinfo {year}
  {2001})}\BibitemShut {NoStop}%
\bibitem [{\citenamefont {Walser}\ \emph {et~al.}(2002)\citenamefont {Walser},
  \citenamefont {Urbach}, \citenamefont {Hatsagortsyan}, \citenamefont {Hu},\
  and\ \citenamefont
  {Keitel}}]{Walser:Urbach:Hatsagortsyan:Hu:Keitel:2002:spin_and_radiation_in_intense_laser_fields}%
  \BibitemOpen
  \bibfield  {author} {\bibinfo {author} {\bibfnamefont {M.~W.}\ \bibnamefont
  {Walser}}, \bibinfo {author} {\bibfnamefont {D.~J.}\ \bibnamefont {Urbach}},
  \bibinfo {author} {\bibfnamefont {K.~Z.}\ \bibnamefont {Hatsagortsyan}},
  \bibinfo {author} {\bibfnamefont {S.~X.}\ \bibnamefont {Hu}},\ \bibnamefont
  {and}\ \bibinfo {author} {\bibfnamefont {C.~H.}\ \bibnamefont {Keitel}},\
  }\href {\doibase 10.1103/PhysRevA.65.043410} {\bibfield  {journal} {\bibinfo
  {journal} {Phys. Rev. A}\ }\textbf {\bibinfo {volume} {65}},\ \bibinfo
  {pages} {043410} (\bibinfo {year} {2002})}\BibitemShut {NoStop}%
\bibitem [{\citenamefont {Hu}\ and\ \citenamefont
  {Keitel}(1999)}]{Hu:Keitel:1999:Spin_Signatures}%
  \BibitemOpen
  \bibfield  {author} {\bibinfo {author} {\bibfnamefont {S.~X.}\ \bibnamefont
  {Hu}}\ \bibnamefont {and}\ \bibinfo {author} {\bibfnamefont {C.~H.}\
  \bibnamefont {Keitel}},\ }\href {\doibase 10.1103/PhysRevLett.83.4709}
  {\bibfield  {journal} {\bibinfo  {journal} {Phys. Rev. Lett.}\ }\textbf
  {\bibinfo {volume} {83}},\ \bibinfo {pages} {4709} (\bibinfo {year}
  {1999})}\BibitemShut {NoStop}%
\bibitem [{\citenamefont {Faisal}\ and\ \citenamefont
  {Bhattacharyya}(2004)}]{Faisal:Bhattacharyya:2004:helium_double_ionization_spin_assymetry}%
  \BibitemOpen
  \bibfield  {author} {\bibinfo {author} {\bibfnamefont {F.~H.~M.}\
  \bibnamefont {Faisal}}\ \bibnamefont {and}\ \bibinfo {author} {\bibfnamefont
  {S.}~\bibnamefont {Bhattacharyya}},\ }\href {\doibase
  10.1103/PhysRevLett.93.053002} {\bibfield  {journal} {\bibinfo  {journal}
  {Phys. Rev. Lett.}\ }\textbf {\bibinfo {volume} {93}},\ \bibinfo {pages}
  {053002} (\bibinfo {year} {2004})}\BibitemShut {NoStop}%
\bibitem [{\citenamefont {Szymanowski}\ \emph {et~al.}(1998)\citenamefont
  {Szymanowski}, \citenamefont {Ta{\"\i}eb},\ and\ \citenamefont
  {Maquet}}]{Szymanowski:etal:1998:laser_scattering}%
  \BibitemOpen
  \bibfield  {author} {\bibinfo {author} {\bibfnamefont {C.}~\bibnamefont
  {Szymanowski}}, \bibinfo {author} {\bibfnamefont {R.}~\bibnamefont
  {Ta{\"\i}eb}},\ \bibnamefont {and}\ \bibinfo {author} {\bibfnamefont
  {A.}~\bibnamefont {Maquet}},\ }\href@noop {} {\bibfield  {journal} {\bibinfo
  {journal} {Laser Phys.}\ }\textbf {\bibinfo {volume} {8}},\ \bibinfo {pages}
  {102} (\bibinfo {year} {1998})}\BibitemShut {NoStop}%
\bibitem [{\citenamefont {Ivanov}\ \emph {et~al.}(2004)\citenamefont {Ivanov},
  \citenamefont {Kotkin},\ and\ \citenamefont
  {Serbo}}]{Ivanov:etal:2004:polarization}%
  \BibitemOpen
  \bibfield  {author} {\bibinfo {author} {\bibfnamefont {D.}~\bibnamefont
  {Ivanov}}, \bibinfo {author} {\bibfnamefont {G.}~\bibnamefont {Kotkin}},\
  \bibnamefont {and}\ \bibinfo {author} {\bibfnamefont {V.}~\bibnamefont
  {Serbo}},\ }\href {\doibase 10.1140/epjc/s2004-01861-x} {\bibfield  {journal}
  {\bibinfo  {journal} {Eur. Phys. J. C}\ }\textbf {\bibinfo {volume} {36}},\
  \bibinfo {pages} {127} (\bibinfo {year} {2004})}\BibitemShut {NoStop}%
\bibitem [{\citenamefont {Ehlotzky}\ \emph {et~al.}(2009)\citenamefont
  {Ehlotzky}, \citenamefont {Krajewska},\ and\ \citenamefont
  {Kami\'nski}}]{Ehlotzky:etal:2009:Fundamental_processes}%
  \BibitemOpen
  \bibfield  {author} {\bibinfo {author} {\bibfnamefont {F.}~\bibnamefont
  {Ehlotzky}}, \bibinfo {author} {\bibfnamefont {K.}~\bibnamefont
  {Krajewska}},\ \bibnamefont {and}\ \bibinfo {author} {\bibfnamefont {J.~Z.}\
  \bibnamefont {Kami\'nski}},\ }\href {\doibase 10.1088/0034-4885/72/4/046401}
  {\bibfield  {journal} {\bibinfo  {journal} {Rep. Prog. Phys.}\ }\textbf
  {\bibinfo {volume} {72}},\ \bibinfo {pages} {046401} (\bibinfo {year}
  {2009})}\BibitemShut {NoStop}%
\bibitem [{Note3()}]{Note3}%
  \BibitemOpen
  \bibinfo {note} {In this article, we use Gauss units and define $\hbar
  =1$.}\BibitemShut {Stop}%
\bibitem [{\citenamefont {Thaller}(1992)}]{Thaller:1992:Dirac}%
  \BibitemOpen
  \bibfield  {author} {\bibinfo {author} {\bibfnamefont {B.}~\bibnamefont
  {Thaller}},\ }\href@noop {} {\emph {\bibinfo {title} {The {Dirac}
  Equation}}},\ Texts and Monographs in Physics\ (\bibinfo  {publisher}
  {Springer},\ \bibinfo {year} {1992})\BibitemShut {NoStop}%
\bibitem [{\citenamefont {Batelaan}(2000)}]{Batelaan:2000:Kapitza-Dirac}%
  \BibitemOpen
  \bibfield  {author} {\bibinfo {author} {\bibfnamefont {H.}~\bibnamefont
  {Batelaan}},\ }\href {\doibase 10.1080/00107510010001220} {\bibfield
  {journal} {\bibinfo  {journal} {Contemp. Phys.}\ }\textbf {\bibinfo {volume}
  {41}},\ \bibinfo {pages} {369} (\bibinfo {year} {2000})}\BibitemShut
  {NoStop}%
\bibitem [{\citenamefont {Braun}\ \emph {et~al.}(1999)\citenamefont {Braun},
  \citenamefont {Su},\ and\ \citenamefont
  {Grobe}}]{Braun:Su:Grobe:1999:FFT_split_operator}%
  \BibitemOpen
  \bibfield  {author} {\bibinfo {author} {\bibfnamefont {J.~W.}\ \bibnamefont
  {Braun}}, \bibinfo {author} {\bibfnamefont {Q.}~\bibnamefont {Su}},\
  \bibnamefont {and}\ \bibinfo {author} {\bibfnamefont {R.}~\bibnamefont
  {Grobe}},\ }\href {\doibase 10.1103/PhysRevA.59.604} {\bibfield  {journal}
  {\bibinfo  {journal} {Phys. Rev. A}\ }\textbf {\bibinfo {volume} {59}},\
  \bibinfo {pages} {604} (\bibinfo {year} {1999})}\BibitemShut {NoStop}%
\bibitem [{\citenamefont {Selst\o{}}\ \emph {et~al.}(2009)\citenamefont
  {Selst\o{}}, \citenamefont {Lindroth},\ and\ \citenamefont
  {Bengtsson}}]{Selsto:Lindroth:Bengtsson:2009:hydrogen_dirac_solution}%
  \BibitemOpen
  \bibfield  {author} {\bibinfo {author} {\bibfnamefont {S.}~\bibnamefont
  {Selst\o{}}}, \bibinfo {author} {\bibfnamefont {E.}~\bibnamefont
  {Lindroth}},\ \bibnamefont {and}\ \bibinfo {author} {\bibfnamefont
  {J.}~\bibnamefont {Bengtsson}},\ }\href {\doibase 10.1103/PhysRevA.79.043418}
  {\bibfield  {journal} {\bibinfo  {journal} {Phys. Rev. A}\ }\textbf {\bibinfo
  {volume} {79}},\ \bibinfo {pages} {043418} (\bibinfo {year}
  {2009})}\BibitemShut {NoStop}%
\bibitem [{\citenamefont {Mocken}\ and\ \citenamefont
  {Keitel}(2008)}]{Mocken:2008:FFT_split_operator}%
  \BibitemOpen
  \bibfield  {author} {\bibinfo {author} {\bibfnamefont {G.~R.}\ \bibnamefont
  {Mocken}}\ \bibnamefont {and}\ \bibinfo {author} {\bibfnamefont {C.~H.}\
  \bibnamefont {Keitel}},\ }\href {\doibase 10.1016/j.cpc.2008.01.042}
  {\bibfield  {journal} {\bibinfo  {journal} {Comp. Phys. Comm.}\ }\textbf
  {\bibinfo {volume} {178}},\ \bibinfo {pages} {868} (\bibinfo {year}
  {2008})}\BibitemShut {NoStop}%
\bibitem [{\citenamefont {Bauke}\ and\ \citenamefont
  {Keitel}(2011)}]{Bauke:Keitel:2011:Accelerating}%
  \BibitemOpen
  \bibfield  {author} {\bibinfo {author} {\bibfnamefont {H.}~\bibnamefont
  {Bauke}}\ \bibnamefont {and}\ \bibinfo {author} {\bibfnamefont {C.~H.}\
  \bibnamefont {Keitel}},\ }\href {\doibase 10.1016/j.cpc.2011.07.003}
  {\bibfield  {journal} {\bibinfo  {journal} {Comp. Phys. Comm.}\ }\textbf
  {\bibinfo {volume} {182}},\ \bibinfo {pages} {2454} (\bibinfo {year}
  {2011})}\BibitemShut {NoStop}%
\bibitem [{\citenamefont {Fillion-Gourdeau}\ \emph {et~al.}(2012)\citenamefont
  {Fillion-Gourdeau}, \citenamefont {Lorin},\ and\ \citenamefont
  {Bandrauk}}]{FillionGourdeau:Lorin:Bandrauk:2012:solution_dirac_equation}%
  \BibitemOpen
  \bibfield  {author} {\bibinfo {author} {\bibfnamefont {F.}~\bibnamefont
  {Fillion-Gourdeau}}, \bibinfo {author} {\bibfnamefont {E.}~\bibnamefont
  {Lorin}},\ \bibnamefont {and}\ \bibinfo {author} {\bibfnamefont {A.~D.}\
  \bibnamefont {Bandrauk}},\ }\href {\doibase 10.1016/j.cpc.2012.02.012}
  {\bibfield  {journal} {\bibinfo  {journal} {Comp. Phys. Comm.}\ }\textbf
  {\bibinfo {volume} {183}},\ \bibinfo {pages} {1403} (\bibinfo {year}
  {2012})}\BibitemShut {NoStop}%
\bibitem [{\citenamefont {Foldy}\ and\ \citenamefont
  {Wouthuysen}(1950)}]{Foldy:Wouthuysen:1950:nr_limit}%
  \BibitemOpen
  \bibfield  {author} {\bibinfo {author} {\bibfnamefont {L.~L.}\ \bibnamefont
  {Foldy}}\ \bibnamefont {and}\ \bibinfo {author} {\bibfnamefont {S.~A.}\
  \bibnamefont {Wouthuysen}},\ }\href {\doibase 10.1103/PhysRev.78.29}
  {\bibfield  {journal} {\bibinfo  {journal} {Phys. Rev.}\ }\textbf {\bibinfo
  {volume} {78}},\ \bibinfo {pages} {29} (\bibinfo {year} {1950})}\BibitemShut
  {NoStop}%
\bibitem [{Note4()}]{Note4}%
  \BibitemOpen
  \bibinfo {note} {An expansion of the wave function into momentum
  eigenfunctions of the Klein-Gordon equation in Hamiltonian form \cite
  {Feshbach:Villars:1954:Relativistic_Wave_Mechanics,*Ruf:2009:Klein-Gordon_Splitoperator}
  yields a system of ordinary differential equations similar to (\ref
  {eq:FSME-dirac-equation}).}\BibitemShut {Stop}%
\bibitem [{\citenamefont {Feshbach}\ and\ \citenamefont
  {Villars}(1958)}]{Feshbach:Villars:1954:Relativistic_Wave_Mechanics}%
  \BibitemOpen
  \bibfield  {author} {\bibinfo {author} {\bibfnamefont {H.}~\bibnamefont
  {Feshbach}}\ \bibnamefont {and}\ \bibinfo {author} {\bibfnamefont
  {F.}~\bibnamefont {Villars}},\ }\href {\doibase 10.1103/RevModPhys.30.24}
  {\bibfield  {journal} {\bibinfo  {journal} {Rev. Mod. Phys.}\ }\textbf
  {\bibinfo {volume} {30}},\ \bibinfo {pages} {24} (\bibinfo {year}
  {1958})}\BibitemShut {NoStop}%
\bibitem [{\citenamefont {Ruf}\ \emph {et~al.}(2009)\citenamefont {Ruf},
  \citenamefont {Bauke},\ and\ \citenamefont
  {Keitel}}]{Ruf:2009:Klein-Gordon_Splitoperator}%
  \BibitemOpen
  \bibfield  {author} {\bibinfo {author} {\bibfnamefont {M.}~\bibnamefont
  {Ruf}}, \bibinfo {author} {\bibfnamefont {H.}~\bibnamefont {Bauke}},\
  \bibnamefont {and}\ \bibinfo {author} {\bibfnamefont {C.~H.}\ \bibnamefont
  {Keitel}},\ }\href {\doibase 10.1016/j.jcp.2009.09.012} {\bibfield  {journal}
  {\bibinfo  {journal} {J. Comp. Phys.}\ }\textbf {\bibinfo {volume} {228}},\
  \bibinfo {pages} {9092} (\bibinfo {year} {2009})}\BibitemShut {NoStop}%
\bibitem [{\citenamefont {Mimura}\ \emph {et~al.}(2010)\citenamefont {Mimura},
  \citenamefont {Handa}, \citenamefont {Kimura}, \citenamefont {Yumoto},
  \citenamefont {Yamakawa}, \citenamefont {Yokoyama}, \citenamefont
  {Matsuyama}, \citenamefont {Inagaki}, \citenamefont {Yamamura}, \citenamefont
  {Sano}, \citenamefont {Tamasaku}, \citenamefont {Nishino}, \citenamefont
  {Yabashi}, \citenamefont {Ishikawa},\ and\ \citenamefont
  {Yamauchi}}]{Mimura:2010:10nm-x-ray-focusing}%
  \BibitemOpen
  \bibfield  {author} {\bibinfo {author} {\bibfnamefont {H.}~\bibnamefont
  {Mimura}}, \bibinfo {author} {\bibfnamefont {S.}~\bibnamefont {Handa}},
  \bibinfo {author} {\bibfnamefont {T.}~\bibnamefont {Kimura}}, \bibinfo
  {author} {\bibfnamefont {H.}~\bibnamefont {Yumoto}}, \bibinfo {author}
  {\bibfnamefont {D.}~\bibnamefont {Yamakawa}}, \bibinfo {author}
  {\bibfnamefont {H.}~\bibnamefont {Yokoyama}}, \bibinfo {author}
  {\bibfnamefont {S.}~\bibnamefont {Matsuyama}}, \bibinfo {author}
  {\bibfnamefont {K.}~\bibnamefont {Inagaki}}, \bibinfo {author} {\bibfnamefont
  {K.}~\bibnamefont {Yamamura}}, \bibinfo {author} {\bibfnamefont
  {Y.}~\bibnamefont {Sano}} \emph {et~al.},\ }\href {\doibase
  10.1038/nphys1457} {\bibfield  {journal} {\bibinfo  {journal} {Nat. Phys.}\
  }\textbf {\bibinfo {volume} {6}},\ \bibinfo {pages} {122} (\bibinfo {year}
  {2010})}\BibitemShut {NoStop}%
\bibitem [{\citenamefont {Zholents}\ and\ \citenamefont
  {Penn}(2005)}]{Zholents:Penn:2005:attosecond_x-ray-pulses}%
  \BibitemOpen
  \bibfield  {author} {\bibinfo {author} {\bibfnamefont {A.~A.}\ \bibnamefont
  {Zholents}}\ \bibnamefont {and}\ \bibinfo {author} {\bibfnamefont
  {G.}~\bibnamefont {Penn}},\ }\href {\doibase 10.1103/PhysRevSTAB.8.050704}
  {\bibfield  {journal} {\bibinfo  {journal} {Phys. Rev. ST Accel. Beams}\
  }\textbf {\bibinfo {volume} {8}},\ \bibinfo {pages} {050704} (\bibinfo {year}
  {2005})}\BibitemShut {NoStop}%
\bibitem [{\citenamefont {Geloni}\ \emph {et~al.}(2011)\citenamefont {Geloni},
  \citenamefont {Kocharyan},\ and\ \citenamefont
  {Saldin}}]{Geloni:Kocharyan:Saldin:2011:self-seeding_X-ray_FELs}%
  \BibitemOpen
  \bibfield  {author} {\bibinfo {author} {\bibfnamefont {G.}~\bibnamefont
  {Geloni}}, \bibinfo {author} {\bibfnamefont {V.}~\bibnamefont {Kocharyan}},\
  \bibnamefont {and}\ \bibinfo {author} {\bibfnamefont {E.}~\bibnamefont
  {Saldin}},\ }\href {\doibase 10.1080/09500340.2011.586473} {\bibfield
  {journal} {\bibinfo  {journal} {J. Mod. Opt.}\ }\textbf {\bibinfo {volume}
  {58}},\ \bibinfo {pages} {1391} (\bibinfo {year} {2011})}\BibitemShut
  {NoStop}%
\bibitem [{\citenamefont {Landau}\ and\ \citenamefont
  {Lifshitz}(1975)}]{Landau:Lifshitz:1975:2}%
  \BibitemOpen
  \bibfield  {author} {\bibinfo {author} {\bibfnamefont {L.~D.}\ \bibnamefont
  {Landau}}\ \bibnamefont {and}\ \bibinfo {author} {\bibfnamefont
  {E.}~\bibnamefont {Lifshitz}},\ }\href@noop {} {\emph {\bibinfo {title} {The
  Classical Theory of Fields}}},\ \bibinfo {edition} {4th}\ ed.\ (\bibinfo
  {publisher} {Butterworth-Heinemann},\ \bibinfo {address} {Oxford},\ \bibinfo
  {year} {1975})\BibitemShut {NoStop}%
\bibitem [{\citenamefont {Keitel}\ \emph {et~al.}(1998)\citenamefont {Keitel},
  \citenamefont {Szymanowski}, \citenamefont {Knight},\ and\ \citenamefont
  {Maquet}}]{Keitel:etal:1998:Radiative_reaction}%
  \BibitemOpen
  \bibfield  {author} {\bibinfo {author} {\bibfnamefont {C.~H.}\ \bibnamefont
  {Keitel}}, \bibinfo {author} {\bibfnamefont {C.}~\bibnamefont {Szymanowski}},
  \bibinfo {author} {\bibfnamefont {P.~L.}\ \bibnamefont {Knight}},\
  \bibnamefont {and}\ \bibinfo {author} {\bibfnamefont {A.}~\bibnamefont
  {Maquet}},\ }\href {\doibase 10.1088/0953-4075/31/3/002} {\bibfield
  {journal} {\bibinfo  {journal} {J. Phys. B: At., Mol. Opt. Phys.}\ }\textbf
  {\bibinfo {volume} {31}},\ \bibinfo {pages} {L75} (\bibinfo {year}
  {1998})}\BibitemShut {NoStop}%
\bibitem [{\citenamefont {{Di Piazza}}\ \emph {et~al.}(2009)\citenamefont {{Di
  Piazza}}, \citenamefont {Hatsagortsyan},\ and\ \citenamefont
  {Keitel}}]{DiPiazza:etal:2009:Radiation_Reaction}%
  \BibitemOpen
  \bibfield  {author} {\bibinfo {author} {\bibfnamefont {A.}~\bibnamefont {{Di
  Piazza}}}, \bibinfo {author} {\bibfnamefont {K.~Z.}\ \bibnamefont
  {Hatsagortsyan}},\ \bibnamefont {and}\ \bibinfo {author} {\bibfnamefont
  {C.~H.}\ \bibnamefont {Keitel}},\ }\href {\doibase
  10.1103/PhysRevLett.102.254802} {\bibfield  {journal} {\bibinfo  {journal}
  {Phys. Rev. Lett.}\ }\textbf {\bibinfo {volume} {102}},\ \bibinfo {pages}
  {254802} (\bibinfo {year} {2009})}\BibitemShut {NoStop}%
\end{thebibliography}%

\end{document}